\documentclass[aps,prl,twocolumn,groupedaddress,showpacs,floatfix,superscriptaddress]{revtex4}
\usepackage{graphicx,epsfig,dcolumn}

\begin{document}

\title{Frequency Metrology on single trapped ions in the weak binding limit:\\The 3$s_{1/2}$--3$p_{3/2}$ transition in $^{24}$Mg$^{+}$}

\author{M. Herrmann}
\author{V. Batteiger}
\author{S. Kn\"unz}
\author{G. Saathoff}
\author{Th. Udem}
\affiliation{Max--Planck--Institut f\"ur Quantenoptik,
Hans--Kopfermann--Strasse 1, 85748 Garching, Germany}
\author{T.W. H\"ansch}
\affiliation{Max--Planck--Institut f\"ur Quantenoptik,
Hans--Kopfermann--Strasse 1, 85748 Garching, Germany}
\affiliation{Ludwig-Maximilians-Universit\"at M\"unchen, 80333
M\"unchen, Germany}

\date{\today}

\begin{abstract}
We demonstrate a method for precision spectroscopy on trapped
ions in the limit of unresolved motional sidebands. By
sympathetic cooling of a chain of crystallized ions we suppress
adverse temperature variations induced by the spectroscopy laser
that usually lead to a distorted line profile and obtain a Voigt
profile with negligible distortions. We applied the method to
measure the absolute frequency of the astrophysically relevant D2
transition in single $^{24}$Mg$^{+}$ ions and find
\mbox{1\,072\,082\,934.33(16)\,MHz}, a nearly 400fold improvement
over previous results. Further, we find the excited state
lifetime to be 3.84(10)\,ns.
\end{abstract}


\maketitle

Virtually all absolute frequency measurements on single trapped
ions reported so far have been performed on narrow transitions in
tightly confined ions, so that the oscillation frequency
$\omega_s$ of the ion around the trap center exceeded the
transition linewidth $\omega_s \gg \Gamma$. In this regime,
called strong-binding limit, the absorption spectrum consists of
a ``carrier'' and a number of motional sidebands separated by
$\omega_s$. Spectroscopy of the carrier eliminates first-order
Doppler- and recoil shifts~\cite{Bergquist1987}, an important
prerequisite for the tremendous accuracies achieved. However, a
variety of interesting transitions exist that can hardly be
studied in this regime due to their large linewidths. In
astrophysics strong dipole transitions observed in quasar
absorption spectra are studied to constrain possible variations
of fundamental constants~\cite{Webb1999}, which creates a demand
for accurate laboratory reference data~\cite{Berengut2006}. In
nuclear physics the structure of halo nuclei is being studied via
isotope shift measurements on dipole
transitions~\cite{Zakova2006, Nakamura2006}, with stringent
requirements on the accuracy that so far could not be met. High
precision spectroscopy outside the strong-binding limit is
challenging since the spectroscopy laser induces
detuning-dependent heating and cooling which distorts the line
profile~\cite{Drullinger1980}. In fact, all previous measurements
in this regime were limited by the conventional spectroscopy
techniques used to overcome the heating effects
(e.g.~\cite{Nakamura2006, Wolf2008}). In this letter we present
both theory and an experimental demonstration of a new
spectroscopy method that essentially removes the limitations due
to the back action of the interrogating laser and allows to
observe a well understood line shape with high signal-to-noise
ratio. The experimental demonstration yields a resonance
statistically indistinguishable from a Voigt profile and thus
allows to determine the line center and -widths with
unprecedented accuracy.

\begin{figure}
\includegraphics[width=\columnwidth]{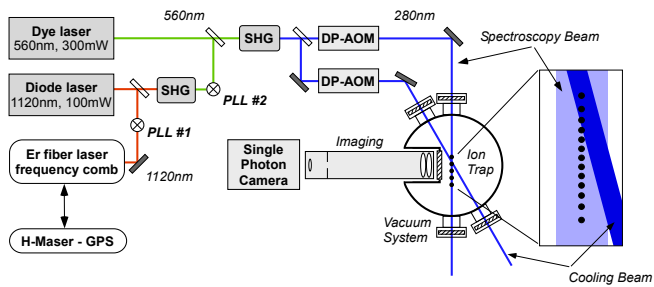}
\caption{Schematic of the optical setup. The dye laser is
phase-locked to a fiber laser frequency comb using a diode laser
as transfer oscillator (phase-locked loops PLL \#1 and \#2). The
output of the dye laser is converted to 280\,nm in a second
harmonic generation (SHG) stage. Spectroscopy and cooling beams
with separately adjustable frequency and stabilized intensity are
obtained by splitting the UV output and passing them through two
double-pass AOM (DP-AOM) setups. The inset shows the geometry of
the cooling and spectroscopy beam relative to the ion chain (to
scale).\label{OpticalSetup}}
\end{figure}

The basic idea is to prepare a crystallized chain of ions stored
in a linear radio frequency (RF) trap which is continuously
laser-cooled on one side only (see Fig.~\ref{OpticalSetup}). Ions
at the other end of the chain are sympathetically cooled but do
not scatter photons from the cooling laser. A spectroscopy beam
less intense than the cooling beam is then directed collinearly
at the ion chain and an imaging photo detector records only
photons from the sympathetically cooled ions. Temperature
variations that lead to line shape distortions are strongly
suppressed, and adverse effects from the cooling laser
(background, ac Stark shift) are eliminated, too.

Measurement schemes involving a dedicated cooling and
spectroscopy beam have previously been used for spectroscopy on
extended samples of ions (Mg$^+$ in a Penning
trap~\cite{Drullinger1980}, Ca$^+$ in a Paul
trap~\cite{Wolf2008}). Unique to our approach is that we perform
spectroscopy on individual ions with spatially separated beams
which, as we will show, enables order-of-magnitude higher
accuracy.

We demonstrate this method by measuring the 42\,MHz wide
3$s_{1/2}$--3$p_{3/2}$ transition near 280\,nm in single
$^{24}$Mg$^+$ ions. This line is an ``anchor line'' for the
many-multiplet method~\cite{Webb1999} used for the search for
drifts of the fine-structure constant in quasar absorption
spectra and has been requested for
re-measurement~\cite{Berengut2006}. Our new approach allows us to
determine the line center to within 160\,kHz, a 375fold
improvement over previous results~\cite{Aldenius2006,
Pickering1998}. Further, this is the first demonstration of an
accuracy better than 1\% of the linewidth in this regime, as
required for the study of halo nuclei~\cite{Zakova2006}. Thanks
to the well understood line shape we could determine the lifetime
of the $3p_{3/2}$ state in excellent agreement with previously
published values~\cite{Ansbacher1989}. The presented measurement
is the first absolute frequency measurement on a single, weakly
bound ion.

\begin{figure}
\includegraphics[width=0.9\columnwidth]{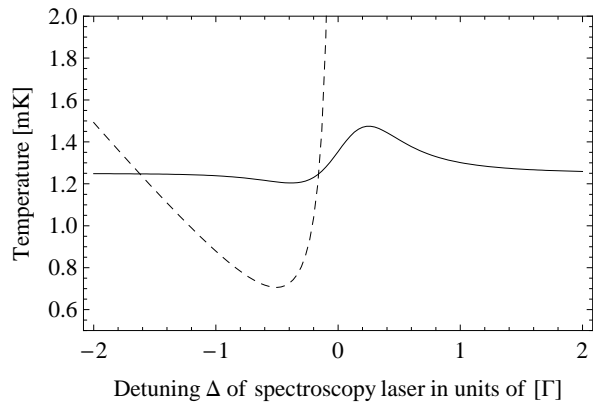}
\caption{Calculated temperature of the spectroscopy ion in a
chain of two ions. The intensity of the cooling laser is set to a
saturation parameter of $s=0.5$ and tuned 64\,MHz below
resonance. The spectroscopy beam is set to $s=0.004$. The dashed
line shows for comparison the temperature profile if the cooling
laser is turned off.\label{comp}}
\end{figure}

To quantify how well our technique can suppress
detuning-dependent temperature variations, consider two
crystallized ions where one ion is cooled while spectroscopy is
performed on the other. The motion can be described as a linear
combination of their two eigenmodes, the center-of-mass and
breathing mode. Since the motional sidebands are not resolved,
the cooling laser will cool both modes simultaneously. For these
conditions we calculate the equilibrium temperature as follows:
The secular cycle-averaged cooling or heating power $\langle
P_l\rangle$ due to the interaction with one of the two laser
beams, enumerated by $i$, reads~\cite{Wineland1979}
\begin{equation}
\langle P^i_{l}\rangle=\langle \hbar k v_0 \cos(\omega_s t) \,
\Gamma g(s_i,\Delta_i-k v_0 \cos (\omega_s t))\rangle.
\end{equation}
Here, $\hbar$ denotes Planck's constant, $k$ the wave vector,
$s_i$ is the dimensionless saturation parameter that measures the
intensity in units of the saturation intensity, $\Delta_i$ the
detuning, $v_0$ the velocity amplitude, $\Gamma$ the natural
linewidth and $g$ finally represents the Lorentzian line shape
$g(s,\Delta)=s/2[1+s+(\frac{2 \Delta}{\Gamma})^2].$ Spontaneous
emission heats the motion on average by
\begin{equation}
\langle P^{i}_{h}\rangle=\langle (1+\xi)\frac{\hbar^2
k^2}{2m}\Gamma g(s_i,\Delta_i-k v_0 \cos (\omega_s t)) \rangle,
\end{equation}
where $m$ is the ion's mass and $\xi=2/5$ for non-isotropic
dipole radiation. Since in both modes the ions move with the same
velocity modulus at any instant of time we can solve the steady
state condition $\sum_i (\langle P^{i}_{l}\rangle + \langle
P^{i}_{h}\rangle)=0$ for $v_0$ to obtain the ion's temperature
$T=m \langle v^2\rangle/2k_b$, where $\langle
v^2\rangle=\frac{1}{2}v_0^2$. The result of such a calculation
with parameters typical for our experiment is shown in
Fig.~\ref{comp}.

This treatment of the cooling dynamics is straight forwardly
generalized to strings of $N$ ions by considering the chain mode
by mode. An important difference is that the cooling laser will
not cool all modes equally efficient if it interacts with a
limited number of ions. For example, the stretch mode of an
odd-numbered string of ions is not cooled if the cooling laser is
focused on the center ion only. An analysis of the normal modes
shows that it is in general favorable to cool a group of ions at
one end of the chain and perform spectroscopy on ions that lie
symmetrically at the other end. For such a configuration,
molecular dynamics simulations show that long ion chains are very
well approximated by chains of two ions as described above if the
intensities $s_i$ are chosen such that the overall ratio of
cooling to heating remains the same. The cooling dynamics remains
unchanged for heterogenous ion chains consisting of different
species.

For a weakly bound ion $\omega_s\ll\Gamma$ in the classical limit
$\hbar \omega_s \ll k_b T$, we expect the line shape to be a
convolution of a power-broadened Lorentzian with a Gaussian due
to the Maxwellian distribution of kinetic
energies~\cite{Stenholm1986}. Thus, for constant temperature a
Voigt profile is expected~\cite{Wineland1979}. However, the
residual temperature variations calculated above cause the line
shape to differ slightly from a Voigt profile. To estimate size
and impact of the deviations we fitted Voigt profiles to
synthetic data generated according to the calculations described
above. For our experimental conditions we find a systematic shift
of the line center of less than 300\,Hz corresponding to
$<10^{-5}$ of the linewidth. The extremely small deviations not
only allow to determine the line center with great accuracy but
also allow to extract information related to the linewidth
precisely, e.g. the temperature of the ion and the lifetime of
the excited state. The departure from a Voigt profile leads to an
error determining the Lorentzian linewidth of less than 0.05\,\%.
These errors can be reduced even further by lowering the
intensity ratio.

To demonstrate the method we have measured the
$3s_{1/2}-3p_{3/2}$ transition near 280\,nm in $^{24}$Mg$^+$.
Ions are stored in a linear RF trap driven at 15.8\,MHz with
secular frequencies of 1\,MHz radially and 60\,kHz axially. The
cooling beam encloses an angle of about $15^\circ$ with the axis
of the ion chain horizontally and $4^\circ$ vertically, limited
by the optical access of our vacuum vessel. The Doppler cooling
limits for the trap's principal axes are thus
\mbox{$T_{\rm{min}}=(0.7,3.5,41.6)$\,mK} along the ion chain,
horizontally and vertically, respectively. The pressure is below
$5\times10^{-11}\,$mbar. We generate the cooling and spectroscopy
beam with separately adjustable frequency and intensity at
280\,nm as follows: The output of a dye laser (Coherent 699/21)
near 560\,nm is frequency-doubled resonantly in
$\beta$-barium-borate (BBO)~\cite{Friedenauer2006} to produce
about 15\,mW at 280\,nm. After spatially filtering the UV beam
with a pinhole it is split and passed through two double-pass
acousto-optic modulators (DP-AOM) that can be tuned within
100$\dots$190\,MHz, allowing us to scan the UV beam over
180\,MHz. The spectroscopy beam is stabilized in intensity and
spatially filtered with an additional 15\,$\mu$m pinhole before
it is focused onto the ions. The focused waist size in the
trapping region is $w_0\cong 200\,\mu$m. For the cooling beam we
use the output of the DP-AOM directly (no intensity stabilization
and spatial filtering) and focus the light on the ions tightly
with a waist of $w_0\cong20\,\mu$m. Both beams are polarized
linearly. To determine the absolute frequency~\cite{Udem2002}, we
phase-lock the dye laser to a 100\,MHz repetition rate erbium
fiber laser frequency comb which is referenced to a
GPS-disciplined hydrogen maser (accuracy $10^{-14}$). A diode
laser serves as transfer oscillator and bridges the gap between
the comb modes at 1120\,nm and the dye laser at 560\,nm. We
determined the mode number of the comb mode by measuring the
frequency of the dye laser simultaneously with an additional
frequency comb operating at a different repetition rate of
250\,MHz. The optical setup is sketched in
Fig.~\ref{OpticalSetup}.

\begin{figure}
\includegraphics[width=0.9\columnwidth]{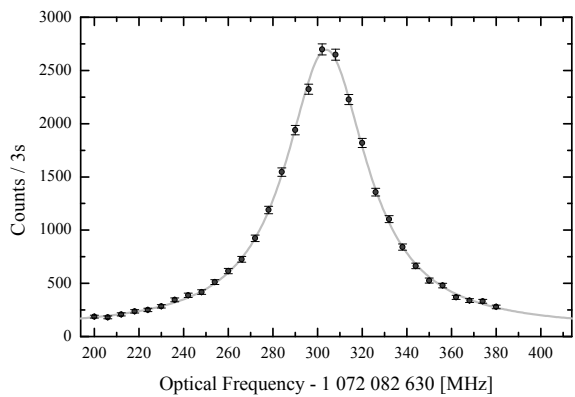}
\caption{Typical recorded resonance. The solid line is a Voigt
fit to the data, the error bars represent Poissonian noise as
derived from the counts.\label{line}}
\end{figure}

To record a line we load chains of 8-12 ions cooled at one end
only by focusing the cooling laser onto 2-3 ions with an
intensity of $s\cong0.5$ and detuned 64\,MHz below resonance (we
did not chose $-\Gamma$/2 for technical reasons). We aligned the
spectroscopy beam collinearly with the axis of the ion chain to
be insensitive to possible micromotion from the radial direction.
Also, this is the direction where cooling is most effective. The
spectroscopy laser is set to a saturation parameter of
$s\cong7\times10^{-4}$. This configuration corresponds to the
parameters given in Fig.~\ref{comp} with six times lower
spectroscopy laser intensity acting on six times more ions. We
collect the fluorescence with an f/2 imaging system capable of
resolving single ions (Quantar Mepsicron II, background count
rate $10^{-4}\,$Hz/pixel, system resolution $2\,\mu$m, total
detection efficiency $6\times10^{-3}$). In one spectroscopy run
we set the spectroscopy laser in random order to 31 different
frequencies in a 180\,MHz broad range centered on the transition.
At each data point we collect photons for 3\,s. We record the
entire image, digitized in 512$\times$512\,pixels and evaluate
the data by selecting circular regions-of-interest around the
three outer most ions. A typical recorded line from a single ion
together with a Voigt profile fitted to the data is shown in
Fig.~\ref{line}. In total, we recorded 264 lines from the outer
three ions in 11 measurement days. The average reduced $\chi^2$
of all lines assuming Poissonian noise only is 1.1. The fit
residuals show no structure other than white noise. Thus, we
cannot detect a statistically significant deviation from a Voigt
profile, in agreement with both our theoretical analysis and the
assumption of shot-noise limited detection.

The high signal-to-noise ratio allows to separate the Lorentzian
and Gaussian contributions to the width and thus to determine the
lifetime of the excited state and the temperature of the ion. The
average widths and statistical uncertainties of the Lorentzian
and Gaussian contribution amount to 41.5(2)\,MHz and
11.5(3)\,MHz, respectively. Residual line shape distortions due
to our method lead to systematic uncertainties of 14\,kHz
(250\,kHz) for the Lorentzian (Gaussian). The low statistical and
method-inherent systematic uncertainties show the potential of
our approach for precision lifetime measurements, corroborated by
excellent agreement with previous
measurements~\cite{Ansbacher1989}. Other systematic uncertainties
not inherent to our method can be larger. With a measured upper
bound on the micromotion modulation index we find from
simulations that the Lorentzian and Gaussian widths are
overestimated by up to 54\,kHz and 170\,kHz, respectively.
Measurements of the polarization dependence of the linewidths
indicate a systematic uncertainty due to magnetic fields of about
1\,MHz. The spectrum of the dye laser available at the time of
the experiment has been measured to be approximately Gaussian
with a linewidth of 5\,MHz and a Lorentzian contribution below
40\,kHz. The width is limited by the servo loops that transfer
some of the frequency comb's short term instability to the laser
despite low feedback bandwidth. The resulting linewidth in the UV
of 10(1)\,MHz gives a residual Doppler width of 6(2)\,MHz. This
corresponds to a temperature of 1.3(8)\,mK, in agreement with the
expected single-ion cooling limit of 1.3\,mK.

\begin{figure}
\includegraphics[width=0.9\columnwidth]{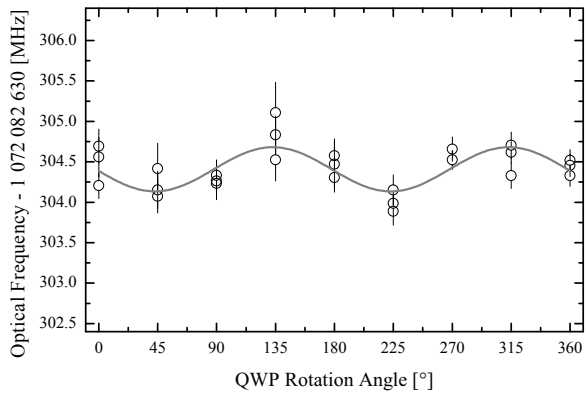}
\caption{Compensation of magnetic field shifts. We measured the
line center versus polarization. The polarization is given in
terms of the rotation angle of the quarter wave plate used;
before the retarder the beam is linearly polarized. For each
polarization three ions were evaluated. \label{qwp}}
\end{figure}

To obtain an accurate line center several systematic
uncertainties need to be taken into account. Static magnetic
fields pointing along the spectroscopy laser shift the line
center due to the linear Zeeman effect if the polarization is not
perfectly linear or optical pumping takes place for other
reasons. The $s_{1/2}$ ground state and the $p_{3/2}$ excited
state have two and four magnetic sublevels, respectively, that
are not resolved for the (laboratory) magnetic fields in our
apparatus. For purely circularly polarized light the line is
shifted by $14\,$kHz/$\mu$T. For this reason we pass the
spectroscopy beam through a quarter wave plate retarder (QWP) and
measure the line center as a function of polarization. If the
polarization is expressed in terms of the rotation angle of the
QWP, a sinusoidal modulation of the line center with a period of
180$^\circ$ is expected. Fig.~\ref{qwp} shows as an example the
line centers of three ions at 9 different angles of the QWP. The
amplitudes of the fitted sines correspond to a magnetic field of
19(3)$\,\mu$T, in agreement with a Hall probe measurement.
Magnetic fields orthogonal to the spectroscopy laser should not
shift the line center, which we confirmed by measuring the line
center with a 200\,$\mu$T field applied perpendicularly.

\begin{figure}[t]
\includegraphics[width=0.9\columnwidth]{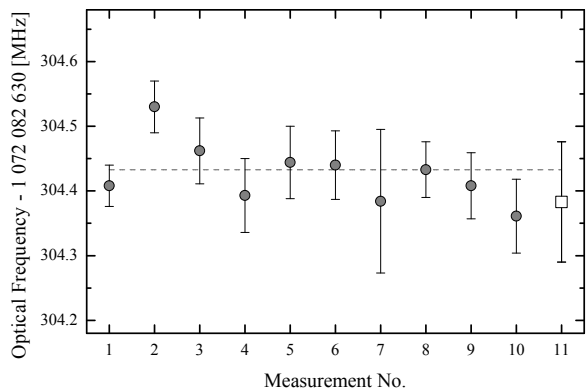}
\caption{Measured line centers. We determined the line center as
in Fig.~\ref{qwp} on 11 days within four weeks. The error bars
represent the statistical uncertainty only. The last data point
was measured with a different laser system (fiber laser).
\label{measdays}}
\end{figure}

The DP-AOMs lead to small detuning-dependent irregular
distortions of the initially Gaussian beam profile. We stabilize
the total intensity but the ions sample the intensity at one
point in space only, so we observe a small alignment-dependent
line center shift which we remove by spatially filtering the
spectroscopy beam with a pinhole. In addition, we evaluate the
data by fitting a Voigt profile including a linearly varying
background to account for residual intensity variations. The
largest systematic uncertainty originates from the large
linewidth of our spectroscopy laser and is not inherent to the
method described here. By phase-locking the laser to the
frequency comb we control the carrier phase. However, correlated
amplitude- and phase-modulation can lead to asymmetries of the
spectrum so its center-of-gravity does no longer coincide with
the carrier frequency. Since the measured line profile is a
convolution of the atomic response with the laser spectrum, this
can lead to systematic shifts. To estimate the size of this
effect we studied the in-loop spectrum of the heterodyne beat
note between the dye laser and the diode laser locked to the
frequency comb. By fitting a Gaussian to the spectrum we
estimated its asymmetry by determining the difference between the
peak of the Gaussian and the counted heterodyne beat signal. We
found an average deviation of 80(60)\,kHz at 560\,nm, so we
assume an uncertainty of 160\,kHz in the UV. In addition we
measured the transition on one day with a different laser system
(a frequency-quadrupled Yb fiber laser~\cite{Friedenauer2006}).
The measurement agreed with our previous measurements (see
Fig.~\ref{measdays}), confirming our estimated uncertainty. Other
systematic uncertainties we considered are significantly smaller:
ac Stark shift due to residual background from the cooling laser
(30\,kHz), dc Stark shift from the trapping fields
(0.1\,Hz)~\cite{Kelleher2008, Sobelman1996}, line shape model
(270\,Hz), maser accuracy (10\,Hz) and 2nd order Doppler shift
(-0.3\,Hz).

To test the reproducibility we repeated the measurements on
eleven days within four weeks. The result is shown in
Fig.~\ref{measdays}.
A weighted fit of the data gives a statistical uncertainty of
15\,kHz, with a reduced $\chi^2$ of 1. Including the correction
due to the recoil shift (-106\,kHz), the absolute frequency reads
\mbox{$\nu=1\,072\,082\,934.33(16)\,$MHz}, in agreement with
\cite{Aldenius2006, Pickering1998} but 375 times more accurate.

We gratefully acknowledge T.~Wilken and R.~Holzwarth for
repairing the fiber laser, B.~Bernhardt for the frequency
dissemination system and H.~A.~Sch{\"u}ssler for assistance with
the dye laser system. This research was supported by the DFG
cluster of excellence ``Munich Centre for Advanced Photonics''.

\end{document}